\documentclass[acmtog, authorversion]{acmart}

\usepackage{booktabs} % For formal tables

\setcitestyle{square}

\usepackage[ruled]{algorithm2e} % For algorithms

\SetAlFnt{\small}
\SetAlCapFnt{\small}
\SetAlCapNameFnt{\small}
\SetAlCapHSkip{0pt}
\IncMargin{-\parindent}

\newcommand{\todo}[1]{\textcolor{red}{\textbf{TODO: #1}}}

\acmJournal{TOG}
\acmVolume{0}
\acmNumber{0}
\acmArticle{0}
\acmYear{2017}
\acmMonth{11}

\acmDOI{0000001.0000001_2}

\begin{document}
% Title portion
\title{Constraint Bubbles: Adding Efficient Zero-Density Bubbles to Incompressible Free Surface Flow}

\author{Ryan Goldade}
\affiliation{
  \institution{University of Waterloo}
}
\email{rgoldade@uwaterloo.ca}
\author{Christopher Batty}
\affiliation{
  \institution{University of Waterloo}
}
\email{christopher.batty@uwaterloo.ca}

\begin{teaserfigure}
    \includegraphics[width=\textwidth]{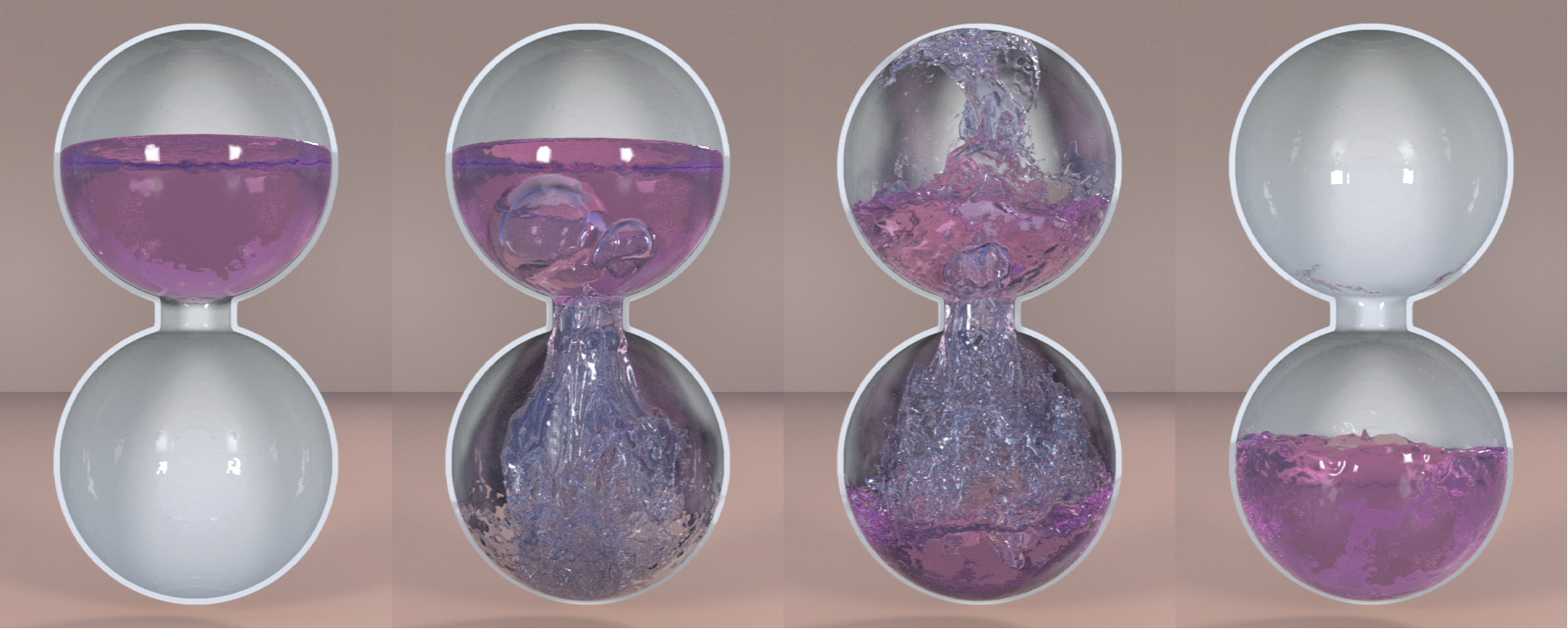}
    \caption{A "water cooler" scenario exhibits a glugging effect, without simulating the air region.
    Our method enforces incompressibility of the bubbles with a single constraint applied over the surface of each air region,
    at only a small additional cost compared to a standard single-phase solver.}
    \label{fig: watercooler}
\end{teaserfigure}

\begin{abstract}
Liquid simulations for computer animation often avoid simulating the air
phase to reduce computational costs and ensure good conditioning of the
linear systems required to enforce incompressibility. However, this
\emph{free surface assumption} leads to an inability to realistically treat
bubbles: submerged gaps in the liquid are interpreted as empty voids that
immediately collapse. To address this shortcoming, we present an efficient,
practical, and conceptually simple approach to augment free surface flows
with negligible density bubbles. Our method adds a new constraint to each
disconnected air region that guarantees zero net flux across its entire
surface, and requires neither simulating both phases nor reformulating into
stream function variables. Implementation of the method requires only minor
modifications to the pressure solve of a standard grid-based fluid solver,
and yields linear systems that remain sparse and symmetric positive
definite. In our evaluations, solving the modified pressure projection
system took no more than 10\% longer than the corresponding free surface
solve. We demonstrate the method's effectiveness and flexibility by
incorporating it into commercial fluid animation software and using it to
generate a variety of dynamic bubble scenarios showcasing glugging effects,
viscous and inviscid bubbles, interactions with irregularly-shaped and
moving solid boundaries, and surface tension effects.
\end{abstract}





\maketitle

\section{Introduction}
The dynamics of submerged air bubbles are critical to the visual realism of
many liquid animation scenarios. However, use of a full-fledged two-phase
flow solver to fully resolve the air dynamics can be problematic for several
reasons: the otherwise unnecessary simulation of the air volume increases the
computational cost; water and air differ in density by about three orders of
magnitude leading to ill-conditioned linear systems that strain standard solvers
(see e.g., \cite{MacLachlan2008}); and the use of a single velocity field for
both liquid and air in a two-phase solver leads to spurious drag effects,
unless treated more carefully \cite{Boyd2012}. Hence, the de facto standard
in computer graphics is to simulate only the liquid region and assume a
\emph{free surface} boundary condition at the liquid-air interface. In other
words, the air is treated as an unsimulated and massless void that has no
influence on the liquid. Unfortunately, doing so has a dramatic and
destructive impact on the observed dynamics: bubbles simply collapse under
the weight of the surrounding liquid, because no force preserves
their volume.

This state of affairs has motivated the pursuit of techniques to add support
for bubbles to free surface flow solvers at lower cost than required for a
tightly coupled two-phase flow. We highlight two relevant approaches. First,
Aanjaneya et al.\ proposed a semi-implicit method for coupling incompressible
liquid to \emph{compressible} bubbles~\cite{Aanjaneya2013,Patkar2013}. While
their tightly coupled compressible-incompressible flow approach is more
complex than even a standard two-phase incompressible flow, the authors also
suggested a constant bubble-pressure simplification that effectively
aggregates the cells comprising a given volumetrically-oscillating bubble
into a kind of super-cell, thereby affording the pressure solve an averaged view of
the bubble. However, since the compressible air mass must still be tracked and
evolved with secondary advection/projection stages, the method nevertheless
scales with the volume of the entire domain rather than that of the liquid alone.
The second relevant approach is that of Ando et al.~\shortcite{Ando2015} who derived a
novel stream function-based discretization with the principal benefit
of supporting genuinely zero-mass bubbles that are incompressible by
construction, without actually simulating them. Unfortunately, this
remarkable property comes at the considerable expense of solving a \emph{vector}
Poisson system that is three times as large as the standard pressure
projection.

Taking our initial inspiration from these two methods, we aim to develop a
straightforward, lightweight, and efficient method to simulate free surface
flows with bubbles, focusing on several key desiderata. First, we aim to
treat bubble regions as massless and completely avoid simulating their
interior air flows. Second, since volumetric oscillations of bubbles are
visually imperceptible in most flows of interest to animation, we favor an
incompressible treatment for simplicity and stability. Third, we prefer a
discretization based on the primitive pressure and velocity variables rather
than stream functions or vorticity, for better compatibility with standard
grid-based free surface flow solvers and the wide variety of extensions that
have been developed to complement them (see e.g., \cite{Bridson2015}).
Fourth, for efficiency we would like the required linear systems to remain
small in size and symmetric positive definite, so as to enable fast solutions
with low memory overhead. While the two methods mentioned above each satisfy some
of these, neither satisfies them all.

Our contribution is therefore a method satisfying all of these goals,
constructed by augmenting a standard free surface flow solver with a
volume-preservation constraint applied to each bubble's boundary. To
illustrate the practicality and efficiency of our method, we implement it
directly inside Houdini's fluid solver \cite{SideEffectsSoftware2017}
and provide performance comparisons with and without bubbles. We
demonstrate our method with a range of bubble scenarios, including rising
bubbles in viscous and inviscid liquid, a glugging water cooler, surface
tension-induced oscillations, and bubble interactions with static and moving
boundaries.

\begin{figure}[h]
	\centering
    \includegraphics[width=\linewidth]{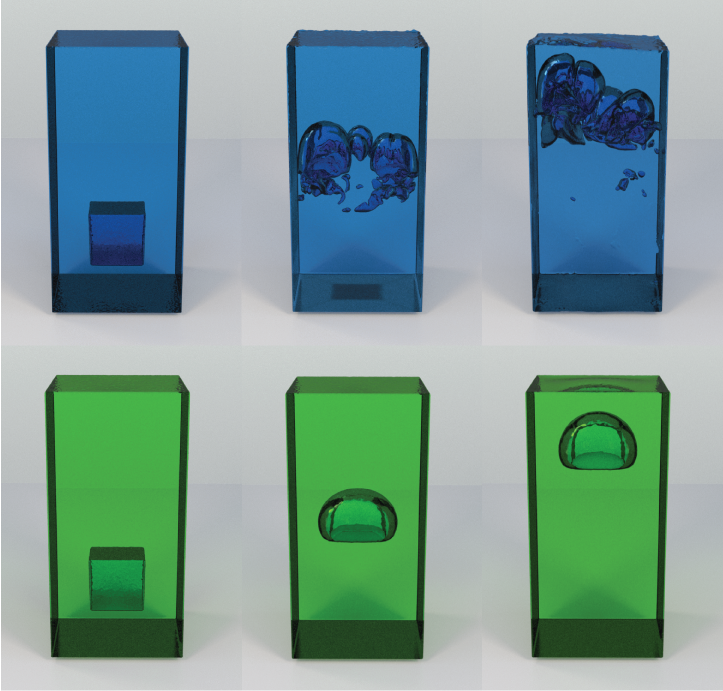}
    \caption{Two columns of fluid with a cubic bubble at the bottom of each, both simulated with our method. The bubble in the inviscid
    simulation (\emph{top}) quickly breaks apart while rising to the
    surface. The viscous simulation (\emph{bottom})
    demonstrates a smooth laminar flow where the single bubble
    remains intact throughout the simulation.}
    \label{fig: risingbubbles}
\end{figure}

\section{Related Work}
We focus our review on the grid-based fluid simulation approaches most
relevant to our work; a useful overview is provided by
Bridson~\shortcite{Bridson2015}. As an alternative, there exist various
smoothed particle hydrodynamic approaches for two-phase flow (e.g.
\cite{Mueller2003,Solenthaler2008}), although they similarly require fully
simulating both materials.

\paragraph{Multiphase Flow}
Many of the approaches used to model two-phase flow in computer graphics
derive from the boundary condition-capturing approach first advocated by Kang
et al.~\shortcite{Kang2000}. This approach simulates both air and fluid,
enforcing incompressibility through a pressure projection scheme that treats
the discontinuous jump in fluid density sharply at the air-water interface
using a ghost fluid method. The first graphics paper to make use of (a
simplification of) this scheme appears to be that of Hong and
Kim~\shortcite{Hong2005}; subsequent variations on this theme include work by
Losasso et al.~\shortcite{Losasso2006} on multiple liquids, Mihalef et
al.~\shortcite{Mihalef2006} on boiling, Kim et al.~\shortcite{Kim2007} on
foams, and Boyd and Bridson~\shortcite{Boyd2012} on FLIP-based two-phase
flow, among others. The work by Kim et al.~\shortcite{Kim2007} is particularly
relevant as it focuses on animating bubbles; however, it differs from our work in
that the air bubbles are all fully simulated, whereas our method avoids this expense entirely.
While two-phase flow approaches typically rely on level set or
particle representations of the interface, coupling with Lagrangian surface
meshes has also been demonstrated \cite{Da2014}. In contrast to these sharp
interface approaches, authors such as Song et al.~\shortcite{Song2005} and
Zheng et al.~\shortcite{Zheng2006} have used a continuous variable-density
pressure solve to simulate multiphase flow, also referred to as
a diffuse interface approach.

\paragraph{Particle Bubbles}
Another natural way to add bubble details to free surface flows is the use of
secondary sub-grid scale particle-based bubbles, coupled in some fashion to
the coarse fluid flow; an early example is the work of Greenwood and
House~\shortcite{Greenwood2004}. More recent instances of this strategy are
the work by Hong et al.~\shortcite{Hong2008a} using SPH bubbles, and that of
Busaryev et al.~\shortcite{Busaryev2012} using weighted Voronoi diagrams of
bubble particles to capture foam structures. An interesting hybrid is the
approach of Patkar et al.~\shortcite{Patkar2013} which essentially unifies
the treatment of sub-grid and grid-scale compressible bubbles to allow tiny
bubbles to both oscillate and coalesce into larger ones.

\paragraph{Augmenting Free Surface Models}
The approaches most related to the current work are those which augment a
free surface flow solver with partially decoupled or fully unsimulated
grid-scale bubbles.

In a computational physics setting, Aanjaneya et al.
\shortcite{Aanjaneya2013} proposed an equation of state approach to simulate
tight two-way coupling of an incompressible liquid to a \emph{compressible}
fully simulated air phase. They also proposed a simplified variant that
assumes constant pressure in the air phase to approximate a bubble's
influence with a single pressure degree of freedom and thereby partially
decouple the air phase. This approach produces a linear system for liquid
incompressibility with a similar structure to ours. However, it involves
extra terms to handle air compressibility and oscillation, and it assumes
that bubbles possess non-negligible air mass that must also be tracked,
necessitating one or more secondary pressure projections over bubble volumes
and conservative advection for the air mass. As such, the method's computational
cost scales with the full domain volume, whereas our method scales with the liquid volume.
The same authors subsequently added sub-grid particle-based compressible bubbles for computer
animation \cite{Patkar2013}. While accurate bubble oscillations are critical to sound
generation (e.g., \cite{Zheng2009}) they are irrelevant for many purely
visual applications, so we instead target a fully incompressible treatment
for zero-mass bubbles, entirely dispensing with the velocity field of the
air.

% should we comment here about our construction being more direct and intuitive like how we argued in our rebuttal?

Ando et al.~\shortcite{Ando2015} proposed a \emph{stream function}-based
approach for free surface flows which expresses the pressure projection
problem in terms of a vector stream function. Standard vector calculus
identities ensure that this representation provides incompressible velocities
for the air by construction, even while assuming the bubbles have zero
density and without simulating air at all. We find their approach very compelling
and believe it is an exciting new avenue of research. However, it is potentially
less attractive in practice for two reasons. First, the
stream function approach entails a radically different and relatively complex
discretization compared to standard solvers, requiring that many existing
solver features, such as surface tension and solid-fluid interaction, be
re-developed from the ground up. Second and more fundamentally, because the stream function is a
three-component vector, the resulting linear systems are vector Poisson
problems three times as large as the usual scalar Poisson problem for
pressure projection, and are therefore significantly slower to solve. The method
we propose instead requires only one extra degree of freedom per bubble and a
small additional computational cost over standard pressure projection.

\paragraph{Constrained Dynamics} Our approach is based on adding extra hard
constraints to a pressure projection solver. The use of Lagrange multipliers
and projection methods for such constraints is longstanding in computer
animation (e.g., \cite{Baraff1996,Goldenthal2007}). They have also been used
in fluid animation for fluid control \cite{Nielsen2010} and for solid-fluid
coupling \cite{Robinson-Mosher2009}; moreover, the pressure itself can be
naturally interpreted as a Lagrange multiplier \cite{Batty2007}.

\section{Smooth Setting}
We base our method on a standard grid-based fluid solver; for further
details, Bridson provides a thorough review \cite{Bridson2015}. In this
context, liquid incompressibility is enforced by the pressure projection step
which performs a projection from the space of all velocity fields onto the
subspace of divergence-free velocity fields. This can be expressed as solving
the PDE
\begin{align}
\begin{split}
\rho \frac{\partial \mathbf{u}}{\partial t} &= -\nabla p, \\
\nabla \cdot \mathbf{u} &= 0. \label{eq:projection}
\end{split}
\end{align}
over the liquid domain, subject to the conditions $p=0$ on free ("air")
surfaces and $ \mathbf{u} \cdot \mathbf{n} = \mathbf{u}_{\text{solid}} \cdot
\mathbf{n}$ at solid walls. In these expressions, $\mathbf{u}$ is the liquid
velocity, $p$ is the fluid pressure enforcing incompressibility, $t$ is time,
$\rho$ is fluid density, and $\mathbf{n}$ is the normal to the solid wall.
\begin{figure}
\centering
    \includegraphics[width=0.75\linewidth]{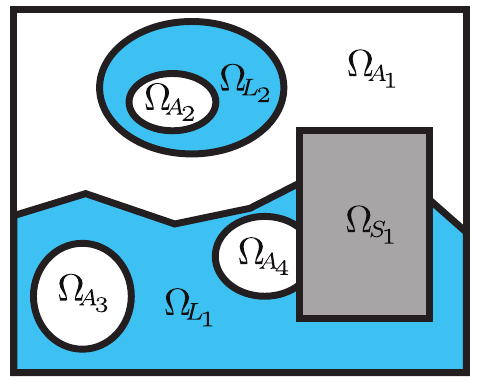}
\caption{The domain of a simulation divided into air (bubble) $\Omega_A$, solid
$\Omega_S$, and liquid $\Omega_L$ regions.}
\label{fig:domains}
\end{figure}

We will replace this step with a new pressure projection augmented with
support for incompressible air bubbles. As shown in Figure \ref{fig:domains},
we divide the simulation volume into three materials identified as air
$\Omega_A$, solids $\Omega_S$, and liquid $\Omega_L$; each of these material
domains may consist of one or more disjoint regions, indicated by integer
subscripts, e.g., $\Omega_{A_1}, \Omega_{L_2}$, etc. We will refer to any
closed disjoint air region as a "bubble". A single connected liquid region
may contain zero or more bubbles within it. A liquid region may also be
entirely surrounded by a single ``bubble"; that is, we make no distinction
between exterior air and submersed air regions, viewing all as bubbles.
Bubble and liquid regions may also be arbitrarily nested. For example, in
Figure \ref{fig:domains}, $\Omega_{A_2}$ is contained by $\Omega_{L_2}$,
 which is itself contained by $\Omega_{A_1}$.

 Our desired behavior is that each bubble should
preserve its total volume. For the $i^{th}$ bubble, we can express this as
the linear velocity constraint
\begin{eqnarray}
\mathsf{B}_i(\mathbf{u}) = \iint_{\partial \Omega_{A_i}} \mathbf{u}\cdot\mathbf{n} dA = 0.
\label{eq:bubbleconstraint}
\end{eqnarray}
That is, the integrated flow through the entire boundary of a single
continuous bubble region, $\Omega_{A_i}$, must be zero. Enforcement of this
constraint involves information about the velocity field everywhere on the
bubble \emph{surface} (i.e., either liquid or solid velocities touching an
air region). Crucially though, no information about velocities
\emph{interior} to the bubble is required.

Collecting all of the bubble constraints, $\mathsf{B}_i$ into a constraint vector
operator $\mathsf{B}$, our PDE becomes
\begin{align}
\begin{split}
\rho \frac{\partial \mathbf{u}}{\partial t} &= -\nabla p - \nabla \mathsf{B}(\mathbf{u})^T\lambda, \\
\nabla \cdot \mathbf{u} &= 0, \\
\mathsf{B}(\mathbf{u}) &= 0,
\label{eq:constrainedprojection}
\end{split}
\end{align}
where $\lambda$ is the vector of Lagrange multipliers having one component
per bubble.

\section{Discretization}
\subsection{Discrete Projection with Bubble Constraints}
We begin by directly discretizing the single-phase PDE \eqref{eq:projection}
on the usual staggered regular grid in finite volume fashion, yielding the
indefinite linear system
\begin{align}
\left(
\begin{matrix}
  M & D^T \\
 D & 0
  \end{matrix}
\right)
\left(
\begin{matrix}
  u \\
  p \\
\end{matrix}
\right)
=
\left(
\begin{matrix}
  Mu^* \\
  0 \\
\end{matrix}
\right).
\label{eq:basicindefinite}
\end{align}
Here $p$ becomes the vector of (discrete) pressures, and $u^*$ and $u$ are
the vectors of velocity face-normal components before and after projection,
respectively. $M$ and $D$ are the usual fluid mass matrix and discrete
divergence operator, which can be straightforwardly adapted to simultaneously
incorporate irregular free surfaces via ghost fluid \cite{Enright2003} and
irregular solid walls via cut-cells \cite{Batty2007,Ng2009}.
Our examples employ this approach. (Note that diagonal entries of $M$ are 0
for entirely air and solid faces, so the corresponding rows and columns drop out.)

We use a row-vector $B_i$ to represent the discretization of the $i^{th}$ bubble
constraint \eqref{eq:bubbleconstraint} which sums
the net flow across the bubble's incident liquid faces such that
\begin{eqnarray}
B_iu = \sum_{\text{liquid faces of $\Omega_{A_i}$}} A_{\text{face}} (\mathbf{u}\cdot \mathbf{n})_{\text{face}} = 0.
\label{eq:discretebubbleconstraint}
\end{eqnarray}
In this expression, $\mathbf{n}$ is the cell face-normal oriented out of the
bubble region, and $A_{\text{face}}$ is the area of the relevant face.
(If a cut-cell methodology is being used \cite{Batty2007,Ng2009}, one should account for only the partial area outside of solids.)
Effectively, this constraint measures the aggregate divergence for the entire
bubble; its corresponding multiplier $\lambda$ will act as a collective
pseudo-pressure enforcing that it be, in total, divergence-free.
Since $B_i$ only involves liquid velocities touching the bubble, it is relatively sparse.

If the bubble touches any kinematically scripted moving solids, we
appropriately modify the right hand side of
\eqref{eq:discretebubbleconstraint} to add contributions from the surfaces of
those solids,
\begin{align}
b_{\text{solid}} = \sum_{\text{solid faces of $\Omega_{A_i}$}} -A_{\text{face}} (\mathbf{u}\cdot
\mathbf{n})_{\text{face}}.
\end{align}
 Doing so allows moving solids to affect even liquid
surfaces that they are \emph{not in direct physical contact with}, such as
when an air bubble in an enclosed tube separates a liquid from a moving
piston: the force is communicated through the bubble, as expected (e.g.
Figure \ref{fig: movingcollision}). The interaction of massless bubbles with
coupled rigid or deformable objects \cite{Batty2007,Robinson-Mosher2009}
could be incorporated in essentially the same manner.

Stacking the bubble constraints into a single fat matrix $B$ and incorporating
them into \eqref{eq:basicindefinite}, we arrive at a large sparse symmetric
indefinite linear system that is the discrete version of \eqref{eq:constrainedprojection}:
\begin{align}
\left(
\begin{matrix}
  M & D^T & B^T \\
  D & 0 & 0 \\
  B & 0 & 0
\end{matrix}
\right)
\left(
\begin{matrix}
  u \\
  p \\
  \lambda
\end{matrix}
\right)
=
\left(
\begin{matrix}
  Mu^* \\
  0 \\
  b_{\text{solid}}
\end{matrix}
\right).
\label{eq:fullsystem}
\end{align}
Unfortunately, the fact that this form includes the velocity degrees of
freedom is troublesome because it substantially inflates the dimensions of
the system. For $N_b$ bubbles and $N_c$ liquid cells (typically with $N_b \ll
N_c$), we have approximately $3N_c$ velocities, $N_c$ pressures, and $N_b$
Lagrange multipliers to solve for, compared to just $N_c$ pressures in the
Poisson problem of the standard bubble-free case.

However, since $M$ is diagonal (i.e., trivially invertible), we can take the
Schur complement to eliminate velocity and arrive at a smaller symmetric
positive definite system in terms of pressure and the bubbles' Lagrange
multipliers:
\begin{align}
\left(
\begin{matrix}
  DM^{-1}D^T & DM^{-1}B^T  \\
  BM^{-1}D^T & BM^{-1}B^T
\end{matrix}
\right)
\left(
\begin{matrix}
  p \\
  \lambda
\end{matrix}
\right)
=
\left(
\begin{matrix}
  Du^* \\
  Bu^* + b_{\text{solid}}
\end{matrix}
\right).
\label{eq:reducedsystem}
\end{align}
Given a solution to this linear system for $p$ and $\lambda$, the velocity
update to recover the final $u$ is given by the first row of
\eqref{eq:fullsystem}. Since $M$ is diagonal, this amounts to a simple
matrix-vector multiply. The upper-left block of \eqref{eq:reducedsystem} is essentially the usual Poisson
system and the remaining blocks account for interaction with the bubble
constraints. We now have $N_b+N_c$ variables compared to the bubble-free case
with $N_c$; that is, we've added one row and column per bubble.

Our system has a similar structure to the one that arises in the compressible
flow method of Aanjaneya et al.~\shortcite{Aanjaneya2013}, but allows for
true zero density bubbles, does not require terms related to bubble expansion
and compression, and incorporates support for moving objects. Furthermore, we
do not require a second advection step or pressure solve to determine the
(visually imperceptible) air motion.

\subsection{Determining Bubble Regions}
Identifying the set of individual bubble regions can be done by determining
connected components through a flooding approach over air cells that share faces.
The flooding must be done over the air volume, rather than just connected surfaces, so
that nesting of regions is properly identified and handled. Our experiments show that
our serial implementation of the flood fill step took approximately 11\% of the
computation time for the pressure solve.
%just clarifying that the flood fill is serial, but the pressure solve was/might not be, right?
.

There are a few situations where we can eliminate one or more of the
bubble constraints, and thereby slightly reduce the size (and potentially
density) of the system. First, if a true free surface effect \emph{is}
desired (e.g., one of the air "bubbles" corresponds to an unbounded exterior
region), then a bubble constraint need not be applied to it,
and the corresponding matrix row and column drop out.

More interestingly, some bubble constraints may be redundant because they are
enforced implicitly by constraints on other incident regions. For example, if
a single incompressible liquid region with a single bubble is completely
contained in a closed solid, the liquid's divergence-free constraint also
ensures that the bubble is divergence-free. That is, since the bubble region
is the geometric complement of an incompressible liquid region, it likewise
cannot expand or compress. If a second disjoint bubble is added to the same
container, a constraint is required on precisely \emph{one} of the two
bubbles; otherwise, one bubble can freely expand while the other contracts to
compensate, despite the liquid remaining divergence-free. In general, for each
volume of space bounded by prescribed-velocity solids and containing any
number of liquid regions and ${n>0}$ bubbles, only ${n-1}$ bubble constraints
are required.

The full set of bubble constraints can alternatively be viewed as introducing a simple null space.
Enforcing all $n$ constraints effectively removes all Dirichlet boundary conditions, leaving only Neumann
boundary conditions along the solid boundary; this is a familiar issue arising
even in single-phase fluid simulation \cite{Bridson2015}.
Instead of resolving this during the linear solve (e.g., \cite{Guendelman2005}), we
remove the null space entirely by deleting a single bubble constraint. This effectively
reintroduces a Dirichlet boundary condition, making the system invertible, and
slightly improves computational efficiency by reducing the number of non-zeros in the linear system.

To maximize the computational benefit of this simplification,
one should first find all mixed fluid-air volumes fully enclosed by solids,
and for each one discard the constraint on the bubble having the largest
\emph{liquid} surface area. Such bubbles lead to coupling among the largest number
of individual fluid pressures, and hence this action corresponds to a reduction
in matrix density by dropping unnecessary rows/columns with the most non-zeros.

% This will need to be updated in the future when we come back with volume control for the
% entire world and we can't just remove a row+column pair from our matrix

\section{Results}
We implemented the proposed method as a direct replacement for the pressure
solve step in Houdini's FLIP solver \cite{SideEffectsSoftware2017}. All of
the examples below were simulated on a six core, i7 5820 CPU. The linear
system \eqref{eq:reducedsystem} was solved using the conjugate gradient method in the Eigen library, using its
diagonal Jacobi preconditioner \cite{Guennebaud2010}.

We found that Houdini's particle resampling tended to suffer from gradual
volume drift, so we applied a simple global liquid volume correction method
in the spirit of Kim et al.~\shortcite{Kim2007}. (By contrast, the stream function
approach \cite{Ando2015} does not support such sources and sinks.)
We emphasize that Kim et al. ~\shortcite{Kim2007} incorporate their divergence terms into
a standard two-phase method \cite{Hong2005} to correct for bubble volume drift. Ultimately, their method
is bound by the computational costs of a two-phase simulation.

% Update by adding a volume control (sub)section that explains how we explicitly track bubbles between
% frames on FLIP particles and how we account for topology changes.

For more accurate bubble surface tracking in the turbulent inviscid rising bubble of Figure \ref{fig:
risingbubbles}, we also sampled the air region with additional
passive particles and used them to correct the surface at each step,
similar to the particle level set method \cite{Enright2003}.

\subsection{Example Scenarios}
\paragraph{Glugging}
Figure \ref{fig: watercooler} demonstrates the familiar glugging effect
exhibited by a \emph{water cooler} scenario. The traditional single phase
free approach surface simply allows the liquid to pour into the bottom bulb
as if both bulbs were open to the outside. By enforcing the bubble
constraints, the downward flow of liquid must match the upward flow of air,
generating a sequence of rising bubbles that are constantly being created and
pinched off.

\paragraph{Rising Bubbles}

In Figure \ref{fig: risingbubbles}, we simulate two initially cubical bubbles
 surrounded by liquid that applies pressure on all sides; the bubble
constraints naturally prevent the liquid from rushing in to fill these gaps.
Instead the vertically increasing pressure in the liquid column forces the
air bubbles upwards, creating the observed rising behavior. The
inviscid example is more turbulent causing the bubbles to break apart and
reconnect. The viscous example exhibits a more laminar flow, maintaining a
single consistent bubble as it rises to the top.

\paragraph{Wall with Holes}

\begin{figure}[h]
	\centering
    \includegraphics[width=\linewidth]{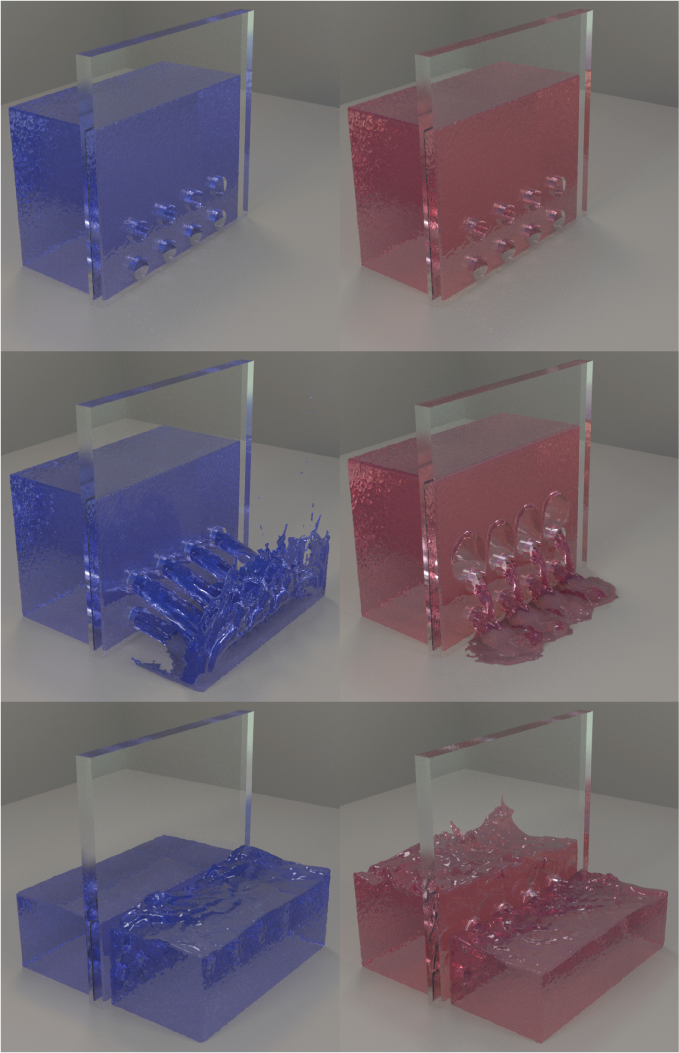}
    \caption{A closed container with a dividing wall in the middle. The
    free surface approach (\emph{left}) allows the liquid to pour through all
    of the holes in the wall unimpeded. Our bubble constraint approach (\emph{right})
    necessitates an opposing flow of air bubbles into
    the left half of the tank, slowing the flow and ultimately preventing the
    liquid levels from equalizing.}
    \label{fig: holywall}
\end{figure}

Figure \ref{fig: holywall} presents an example of a completely enclosed
container with a dividing wall in the middle. The divider contains two rows
of holes to allow the liquid to flow through into the initially empty side.
The fluid in the free surface example flows rapidly and continuously through
both rows of holes until the fluid level is equal on both sides of the wall.
By contrast, with our constrained bubble model the fluid only flows
continuously through the bottom row, because the liquid flow must be balanced
by air bubbles flowing through the top row in the opposing direction. In
addition, the holes soon become fully submerged, which prevents any further
air from passing through the holes. This in turn prevents the liquid levels
from equalizing because the volume of air on both sides of the wall can no
longer change.

\paragraph{Moving Boundary}

\begin{figure}[h]
	\centering
    \includegraphics[trim={0 23cm 0 0},clip,width=\linewidth]{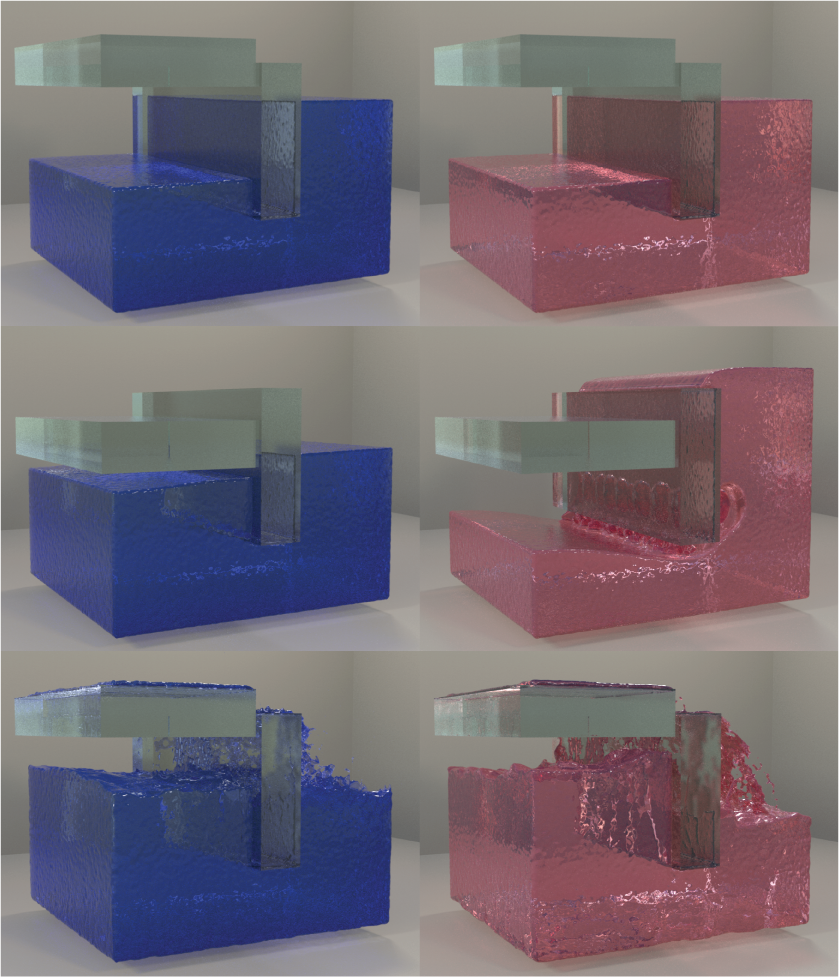}
    \caption{A closed container with a partial vertical wall in the middle and a moving solid platform on the left.
    The free surface approach (\emph{left}) erroneously allows the liquid to rise until it collides with the descending platform (second row).
    Our approach (\emph{right}) instead enforces the air pocket's incompressibility, so the platform pushes
    both air and liquid downwards, creating bubbles and forcing the liquid to rise on the \emph{right} side of the divider.}
    \label{fig: movingcollision}
\end{figure}

Figure \ref{fig: movingcollision} demonstrates how moving solid boundaries
interact with our incompressible air constraint. When the solid boundary
moves down, it creates a net flux at the solid-air boundary that must be
compensated by an opposing flux at the air-liquid boundary. As a result the
air is pushed down and under the dividing wall to create a row of rising
bubbles, at which point additional liquid can enter and begin filling up the
space underneath the solid boundary. At the end of the simulation, the liquid
levels again remain imbalanced because of the incompressible air volume
trapped underneath the solid platform. In the single phase case, we see
physically incorrect behavior: the liquid level immediately begins rising
under the moving solid and equalizes at the end.

\paragraph{Surface Tension}

\begin{figure}[h]
	\centering
    \includegraphics[width=\linewidth]{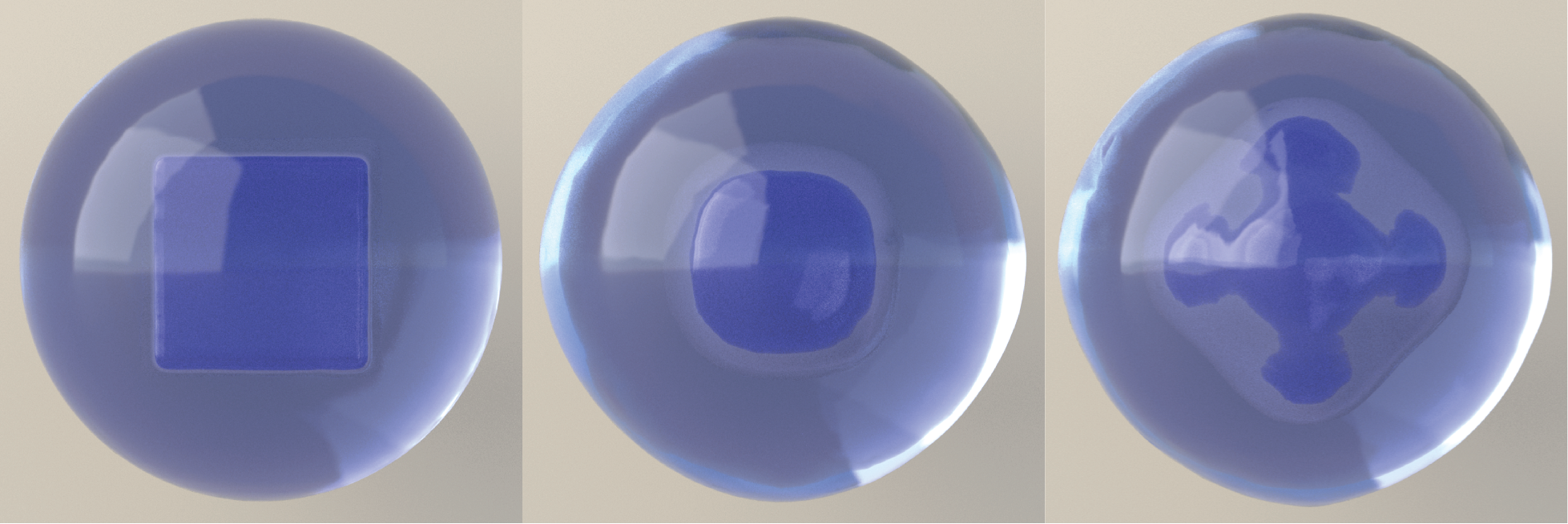}
    \caption{A cube-shaped bubble oscillates inside a liquid sphere in zero gravity.}
    \label{fig: surfacetension}
\end{figure}

Surface tension effects are easily incorporated into our new
bubble-constrained pressure projection. We use the standard ghost fluid
method to apply the surface pressure jump across the liquid surface
\cite{Enright2003,Hong2005}, exactly as in a regular free surface solve.
Figure \ref{fig: surfacetension} presents an initial cube-shaped bubble
inside a sphere that oscillates due to the surface tension forces on the
liquid, and because of incompressibility it does not collapse.

\subsection{Performance Evaluation}
A possible shortcoming of the formulation given by \eqref{eq:reducedsystem}
is that the row and column corresponding to a given bubble can be relatively dense
depending on the bubble's surface area. This is because each bubble
constraint \eqref{eq:discretebubbleconstraint} involves \emph{all} liquid face
velocities incident on that bubble; elimination of the velocity variables
leads to coupling between the bubble's Lagrange multiplier and the pressures
of all its incident cells. This adds some overhead compared to a pure free
surface flow solver (in addition to the cost of identifying bubble regions).

\begin{table*}[th]
	\centering
    \renewcommand{\arraystretch}{1.2}% Wider
	\begin{tabular}{ | l | c | c | c | c | c | }
\hline
	Initial Condition & Grid Resolution & Particle Count & Simulation Time & Max Substeps & \# of Frames \\
\hline
    Water cooler  & 128$\times$256$\times$128 & 4M & 2h55m & 5 & 300 \\
    Wall with holes & 128$\times$128$\times$128 & 4.8M & 2h29m & 5 & 300 \\
    Rising bubbles (inviscid) & 128$\times$256$\times$128 & 17.8M & 7h30m & 5 & 130 \\
    Rising bubbles (viscous) & 128$\times$256$\times$128 & 17.8M & 8h53m & 5 & 180 \\
	Moving solid & 150$\times$150$\times$150 & 6.6M & 4h20m & 5 & 300 \\
    Surface tension & 200$\times$200$\times$200 & 4.1M & 3h28s & 20 & 300 \\
    \hline
	\end{tabular}
	\caption{Computational costs and data for several our various bubble scenarios.
   (An additional 3.8 million passive air particles were used to improve surface tracking for the inviscid rising bubble. The surface tension example averaged four particles / voxel, whereas all other examples averaged eight particles / voxel.)}
	\label{tab: timings}
\vspace{-15pt}
\end{table*}

We explored the impact of this overhead by examining the first ten frames of
both the water cooler and moving solid examples. In these early frames, the
geometry of the liquid is very similar between the two solver methods which
provides a reasonable point of comparison. We examine only the pressure solve
step, since the remaining solver components are untouched. To provide a fair
comparison, pressure solves for the free surface simulations were performed
by simply disabling the bubble feature in our solver code, rather than, for
example, using Houdini's more optimized internal pressure solver.

For the water cooler example, the free surface method required 41 substeps
(about 4 per frame), taking a total of 54.2 seconds for the pressure
projection step, or 1.3 seconds per substep. Our approach required only 26
substeps (2 or 3 per frame), taking a total of 36.4 seconds for our solver,
or 1.4 seconds per substep. The free surface solver required an average of
245 conjugate gradient iterations to converge to a relative residual error of
$10^{-5}$, while our approach required an average of 263 iterations. We make
two observations here. First, our approach requires slightly more iterations
and slightly more time to solve \emph{per substep}.  Second, in this
particular scenario, our approach actually took less time \emph{in total},
because bubbles prevent the liquid from rushing through the neck of the water
cooler with the high velocity seen in the free surface case; thus fewer substeps
were needed to satisfy a reasonable CFL number

We also investigated the moving solid simulation, in part because it requires
the small additional cost of integrating over the solid surface to build the
right-hand side of (\ref{eq:reducedsystem}). The free surface method required
20 substeps over ten frames, taking a total of 1m12s for the pressure
projection step, or 3.6s per substep. Our approach required 41 substeps,
taking a total of 2m38s for our solver, or 3.9s per substep. The solvers
averaged 405 and 411 iterations, respectively. Hence our method was again
just slightly more expensive \emph{per step}. However, in this case it was
slower overall due to the greater number of substeps. This higher substep
count occurred because the liquid moves more quickly when being impulsively
pushed by the air incident on the moving solid, compared to gradually
accelerating under gravity in the free surface case.

In all four of the simulation settings above, the pressure solve took
approximately half of the computation time per substep (the remainder was
spent on Houdini's advection, reseeding, APIC velocity transfer, etc.) For
completeness, the total computation times for all of our bubble simulations
(including additional steps like advection) are presented in Table \ref{tab:
timings}.

In summary, we observed that the components modified by our bubble approach
are typically no more than 10\% slower and require only slightly more
iterations \emph{per substep} compared to the standard free surface pressure
projection, even when handling moving boundaries. However, because the
presence of bubbles often dramatically changes the fluid velocities and
resulting motion, it is not possible to make a truly general statement about
the relative \emph{total costs} other than to state that they are often broadly
similar. For example, the entire water cooler simulation with bubbles took
2h55m while the single-phase version took 3h05m, although their behavior was
radically different. Nevertheless, we are confident in claiming that our
method will be substantially more efficient than either solving for the
entire air side velocity field \cite{Hong2005,Boyd2012,Patkar2013} or using
the stream function formulation \cite{Ando2015} that entails solving a vector Poisson system
with three times as many degrees of freedom.

\begin{comment}
\todo{NB: I simulated at 60fps but played back at 30 fps.. I just thought the
results looked better this way. I don't know if it's worth mentioning
anywhere though}
\end{comment}

\section{Discussion and Conclusions}
We have presented a simple and efficient strategy to add zero-density bubbles
to grid-based liquid solvers, without adopting stream functions or simulating
the air volume. We believe it can be readily adopted into existing tools, as
we have demonstrated for Houdini. In fact, our proposed approach shipped as a new
feature in Houdini 16.5.

Because we do not strictly track per-bubble volume, gradual bubble volume drift and
occasional loss of small bubbles can occur, as with prior multiphase approaches.
If precise preservation is critical, several
treatments are possible. Absent bubble topology changes, the approach of Kim
et al.~\shortcite{Kim2007} can compensate drift with per-bubble divergence
sources. In the presence of topology changes, bubble "rest volumes" are no
longer constant, which necessitates updating them after each topology
change; while easily done for explicit meshes \cite{Thuerey2010}, this is
non-trivial for implicit surfaces. A more costly implicit alternative is to
track air mass as a scalar and fully solve the air field's motion,
conservatively advecting the air mass \cite{Aanjaneya2013}.

Solving for the air field itself may still be desired for certain effects, e.g.,
smoke-filled bubbles rising through liquid. Losasso et
al.~\shortcite{Losasso2006} suggested an efficient decoupled approach,
solving the liquid first using free surface conditions, and then the air
using the liquid velocities as boundary conditions. However, the free surface
condition will again lead to collapsing bubbles; replacing the first solve
with our method will maintain bubbles and ensure compatible boundary
conditions for the air solve.

Another natural extension would be to couple with dynamic rigid and
deformable bodies by accounting for their surface velocities in the integral
for the air bubble incompressibility constraint. Further generalizing the
linear boundary constraint may enable other interesting control effects, such as
bubbles with constraints on translational velocity, rotational velocity, or
shape.

\begin{comment}
\todo{This is just for brainstorming, likely want to remove it.} Interesting
possible extensions:
\begin{enumerate}
\item Control: constraint liquid region to have desired local vorticity, or
    total vorticity magnitude, etc.
\item Control: What other linear velocity constraints would be practically
    useful, and not redundant wrt standard solid boundaries?

\end{enumerate}
\end{comment}
\begin{comment}
\begin{acks}
\end{acks}
\end{comment}

% Bibliography
\bibliographystyle{ACM-Reference-Format}
\bibliography{bubbles}

\end{document}